\documentclass[12pt]{article}
\usepackage{fleqn,espcrc1,epsfig,psfig}
\usepackage{graphicx}
\renewcommand{\baselinestretch}{1.1}

\begin{document}
\setcounter{page}{0}
\thispagestyle{empty}
%
%
%
\markboth{ }{ }
\renewcommand{\baselinestretch}{1.0}\normalsize

\vspace*{2cm}
\renewcommand{\thefootnote}{\fnsymbol{footnote}}
\begin{center}
{\LARGE\bf Evolving QCD\footnote{Supported by the European Network on Electron Scattering off 
Confined Partons (ESOP) under\\ contract no. HPRN-CT-2000-00130 }}
\end{center}
\bigskip
\makeatletter \begin{center}
\large H.J.~Pirner  \vspace*{0.3cm} \\
{\it Institut f\"ur Theoretische Physik der Universit\"at Heidelberg }
\end{center} \makeatother
{\begin{center} (\today) \end{center}}
\vspace*{2cm}
\begin{abstract}
\noindent

We give an overview of the current theoretical status
of proper time renormalization group flow equations
applied to QCD. These equations give the evolution of 
coupling constants in an effective QCD Lagrangian as a function
of an infra-red cut-off parameter. This parameter characterizes the
resolution with which the system is looked at.
Decreasing resolution 
transforms partonic quarks into constituent quarks
and generates pion bound states.
The evolution equations can also be applied 
to finite temperature and finite density QCD.

\end{abstract}

\newpage
\renewcommand{\baselinestretch}{1.1}\normalsize
\renewcommand{\thefigure}{\arabic{figure}}
\renewcommand{\thefootnote}{\arabic{footnote}}
\setcounter{footnote}{0}

%
%
%
\setcounter{section}{0}

Quarks and gluons are the ultimate building blocks of hadrons.
Their interaction is governed by the QCD
coupling which grows dynamically for low energies.
With the increasing coupling gluons and quark-antiquark
pairs multiply indefinitely. Therefore 
QCD forms new structures of dominant collective
units in the infra-red. 
The renormalization group is the principal tool to investigate the 
formation of these new structures. Its importance in quantum field theory 
can only be compared to  the Schr\"odinger equation in quantum mechanics.
In this paper I use the term renormalization group in a wide sense
including  the evolution of systems of Lagrangians.
It may well be that the degrees of freedom change under evolution.
Such an approach is not so well researched yet, but the  
necessity of a theoretical framework for it - especially in QCD -
is universally recognized.  
I will  try to give a summary of our work  
using renormalization group techniques to
handle quark degrees of freedom evolving into hadronic  bound states.

The quark-hadron transition governs
strong interaction physics  when the resolution is lowered below $1 $ GeV.
Hadronic physics  
is well researched up to a mass scale of $1.5$ GeV. 
Its spectroscopy still lacks a clear  identification
of glueballs and hybrid states. 
Yukawa models like the sigma model are natural to
describe hadron physics at low energies.
In order to bridge the gap between quark and meson physics we
choose a hybrid approach where the original action contains a 
four-quark interaction which
is bosonized into sigma and pion fields.
These mesonic degrees of freedom are not propagating fields 
in the ultraviolet, i.e., their kinetic term is zero. They
couple to the quark fields and mimic the local four-fermion
interaction. 
In principle all the mesonic couplings arising from the Fierz 
transformation of a local color exchange interaction should be
taken into account. Mean field theory shows a 
basic ambiguity to the possible Fierz transformations 
\cite{Jaeckel:2002rm}. The RG-equations can overcome this problem.
At the ultraviolet scale the meson-quark 
Yukawa coupling can be set to unity. There is only
the mass parameter of the combined sigma and pion fields in the
Wigner mode which sets the mass scale for the evolving system.
The renormalized Yukawa coupling decreases strongly in the infra-red
which makes the evolution infra-red stable. In the approximation
with meson loops we have indications that the infra-red evolution has
fixed point character. Reversely the  divergence 
of the Yukawa coupling for large scales 
signals the unsatisfactory high energy behaviour of the 
hadronic theory. Above the $1$ GeV scale the compositeness of 
hadronic objects becomes important and one has to choose a
quark-gluon basis.

\begin{figure}[h!]
\setlength{\unitlength}{1.cm}
  \begin{center}
\epsfig{file=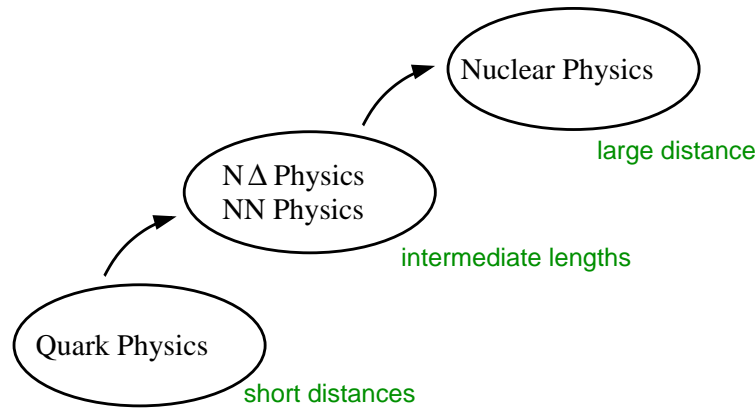, width=10cm} 
  \end{center}
  \caption{The traditional separation of quark physics, 
the physics of the nucleon/delta and nucleon-nucleon interaction 
and nuclear physics}
  \label{fig:Trad}
\end{figure}

In Fig. 1 we summarize the traditional approach to hadronic and nuclear
physics indicating the
hier\-archy between quark 
physics, nucleon-mesonic physics and nuclear physics. A division into
these subfields is highly efficient when special topics in a field 
are researched, because
then refined methods can be applied to get maximum insight
into a detailed aspect. On the other hand it is refreshing to 
cross the borderlines of these fields 
and see how things are connected. Indeed modern  high energy experiments 
in  nuclear physics look at nuclei with
high resolution in electron nucleus collisions  
and  at high excitation energies in
relativistic heavy ion collisions. They naturally relate nuclear
physics to quark physics.

The renormalization group connects regions of different
length scales and is the appropriate tool for such an interdisciplinary
approach. It avoids the appearance of arbitrary cut-off functions when
additional quantum corrections appear. Going beyond the
large $N_c$ appro\-xi\-mation in the Nambu Jona Lasinio (NJL) model one needs to
include meson loops which necessitates an additional cut-off
in the Schwinger-Dyson approach. The unifying aspect of the renormalization
group is shown in Fig. 2 and illustrated by the work  here.
Starting from the renormalization group in the vacuum, which only tests
physics along the resolution axis, I will cover high temperature 
physics and high density investigations in the same theoretical 
framework.

\begin{figure}[h!]
\setlength{\unitlength}{1.cm}
  \begin{center}
\epsfig{file=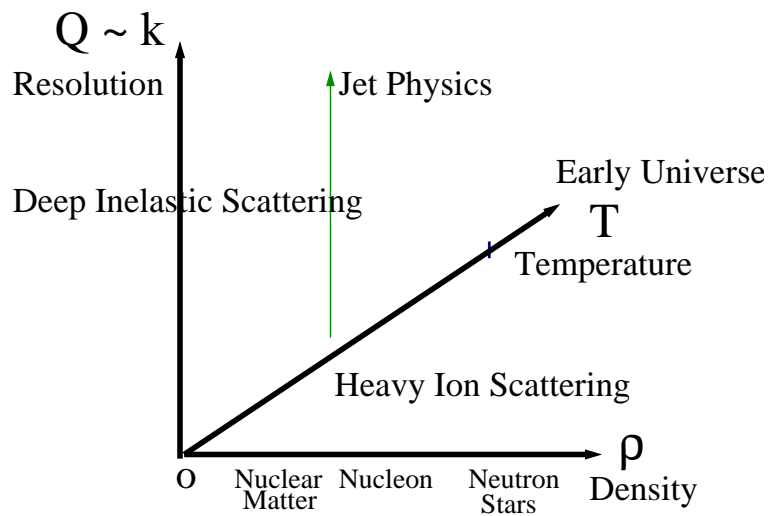, width=10cm} 
  \end{center}
  \caption{The renormalization group unifies physics along
the three different axis: physics at varying resolution, finite
temperature
and density}
  \label{fig:Trad}
\end{figure}

Renormalization group  flow equations describe the average of an
effective action and represent the continuum analogue of a block spin
transformation.  I have tried before to use the Monte Carlo 
renormalization group to extend the lattice gluon action of Wilson into
the long distance region \cite{Pirner:1991im,Pirner:1987qj,Grossmann:1990ws}. 
This approach introduces new collective
variables into the lattice action, namely the colordielectric fields.
The resulting QCD action has a strong 
coupling which is modulated by the colordielectric 
field and confines. It 
reproduces results obtained on the finer lattice with an accuracy
of $10$ \%. A large numerical effort is needed to solve a set of
overdetermined Schwinger-Dyson equations. Already at that time it
was desirable to have an analytic scheme to follow the flow of
couplings when  high momenta larger than a
coarse-graining scale $k$ are integrated out. The flow equations 
proposed by Wegner and Houghton \cite{Wegner}, Polchinsky \cite
{Polchinski}, Wetterich and  Berges \cite{Wetterich,Berges:2000ew} 
present such a scheme. They are 
ultraviolet and infra-red finite through the introduction of a scale
dependent cut-off function. 
We use renormalization group flow equations with a heat kernel cut-off
\cite{Schaefer:1998my,Schaefer:em,Papp:1999he} which we think are
more practical. Moreover, they are so simple that they are 
ideally suited to learn and 
teach renormalization in field theory. 
No infinities appear at any time.
All integrals can be done analytically and have a simple form.
The solution of these RG-equations
includes complicated summations of diagrams which are 
similar to solutions of a coupled set of Schwinger-Dyson equations.
The heat kernel regularization
method \cite{Ball} preserves the symmetries of the theory
and keeps the physical interpretation of the evolution equations
particularly simple for phenomenological applications.

We choose as a starting action at the UV scale the NJL Lagrangian

\begin{equation}
\label{eq:lagrangian-njl}
S_{\mbox{\tiny NJL}} =\int d^4 x  [ \bar{q} \, i{\gamma \partial} q 
+ G \;\left\{(\bar{q} q)^2 + (\bar{q} 
i\vec{\tau}\gamma_5 q)^2 \right\}]\; 
\end{equation}
which is embedded into a linear $\sigma$-model with $Z_q=$ $1$, a 
vanishing wave
function renormalization constant $Z_\Phi=0$ and   
 an effective potential $U(\Phi^2)= \frac{m^2 \Phi^2}{2}$.
We assume the linear $\sigma$-model to be a valid description of nature
below scales of $1.5 $ GeV. In this region
the gluon degrees of freedom are supposed to be already frozen out.
Confinement is not dealt with correctly. The physics is governed by chiral
symmetry, which is spontaneously broken in the infra-red. The Euclidian
action of the linear $\sigma$-model in generalized form is
given by

\begin{equation}
  \label{eq:uv-action}
  S[\Phi,\bar q, q] = \int \! d^4x \left( Z_q(\Phi^2) \bar{q}{
  \gamma\partial} q +\frac{1}{2} Z_{\Phi} (\Phi^2 ) \left(
  \partial_\mu \Phi^a \right) \left( \partial_\mu \Phi^b \right) +
  g(\Phi^2) \bar{q} M q + U(\Phi^2) \right). 
\end{equation}

The quark and meson fields are $\bar q, q$ and $ \Phi$, 
where $\Phi=(\sigma,\vec \pi)$ is the $O(4)$-
and $M=\sigma+i\vec\tau \vec\pi \gamma_5$ the chiral $SU(2)_L \otimes
SU(2)_R$-representation of the meson fields.
We consider this  action to be preserved during the
evolution and follow the generalized couplings and the effective potential
during evolution by calculating the effective action 
$S_{eff}$ in a one-loop approximation. 

\begin{equation} 
S_{eff}[\Phi,\bar q, q] = S[\Phi,\bar q, q] - \frac{1}{2} {\rm Tr} \,
\log S_{\bar q q} + \frac{1}{2} {\rm Tr}
\, \log \left( S_{\Phi^i \Phi^j} - 2 S_{\Phi^i q} S_{\bar q
    q}^{-1} S_{\bar q \Phi^j} \right) \, ,
\end{equation}
with
\begin{eqnarray}
  S_{\bar q q}(x,y)& =& \left. \frac{\delta^2 S[\Phi,\bar
  q, q]}{\delta \bar q(x) \delta q(y)} \right|_{av} \quad , \\
 S_{ \Phi^i \Phi^j}(x,y)&=&\left. \frac{\delta^2 S[\Phi,\bar q,
  q]}{\delta \Phi^i(x) \delta \Phi^j(y)} \right|_{av} .
\end{eqnarray}

The first logarithm results from the fermion loop fluctuations,
whereas the two terms in the second logarithm are the contributions of
the bosonic and the mixed loop, respectively. 
The derivative of $S_{eff}$ with respect to the evolution scale 
can be computed after
a cut-off function is introduced.
For this purpose we represent the fluctuation determinants by Schwinger
proper time integrals with the  cut-off function $f(k^2 \tau)=
\left(1 + k^2\tau + \frac{1}{2}(k^2\tau)^2 \right) \;
  {\mathrm e}^{-k^2\tau}$

\begin{equation}
{\rm Tr}\  log(A)  = -{\rm Tr}\int_0^\infty \frac{d \tau}{\tau} e^{-\tau A} 
f(k^2
\tau).
\end{equation}

The heat kernel cut-off function $f$ suppresses all fluctuations with
momenta below the cut-off scale $k$. Going to $k\rightarrow 0$ means to
include more and more infra-red modes. Finally, all modes are included,
because of the limiting behaviour $f_{k\to 0} \rightarrow 1$. 
The ultraviolet region is left undisturbed due to the
extra $\tau$ and $\tau^2$ terms in front of the exponential. 
Therefore we can extend the $\tau$-integration to the
interval $[0,\infty]$. 
We further replace the masses and couplings of the classical action 
 by the running 
masses and
running couplings of the {\em effective\/} action
$S_{eff}$. Generically, we obtain the following evolution equation
for $S_{eff}$ which contains its second order derivatives $S_{eff}''$
indicated in Eqs.(4,5):
\begin{equation}
\frac{\partial S_{eff}}{\partial k}=-\frac{1}{2} {\rm Tr}
\int \frac{d\tau}{\tau} e^{-S_{eff}^{''}(k)} \frac{\partial}{\partial k}
f(k^2 \tau).
\end{equation}

This replacement of $S''$  by $S_{eff}''$ turns the
one-loop equation into a renormalization group improved flow equation,
which goes beyond the standard one-loop renormalization group running
and includes higher loop terms successively into the proper time
integral. Therefore it is capable to include nonperturbative
physics in the strong coupling region. The evolution equations are 
obtained by comparing the derivative on $S_{eff}(k)$ with respect to $k$ 
with the formal expression of Eq.~(2) which contains  the generalized
couplings and the effective potential as functions of $k$. The local
potential terms are easy to evaluate and lead to nonlinear partial 
differential equations containing derivatives of the potential with 
respect to $\Phi^2$. The evolution of the other  couplings can be
obtained in a derivative expansion which is an expansion in 
derivatives (momentum) over mass. This expansion is well controlled
because
of two reasons. Firstly,  we approach the infra-red, so momenta become always
smaller, we need only few terms in the expansion. Secondly,  the infra-red
cut-off scale $k$ acts like a mass cut-off, so the expansion is well defined
even if the excitations themselves are massless, like the quarks 
in the ultraviolet or the mesons in the infra-red region.

Let me give an overview of  the results  
we have achieved so far.
We have solved in local potential approximation the vacuum evolution,
the finite temperature evolution  
and finite baryon density evolution 
\cite{Schaefer:1998my,Schaefer:em,Papp:1999he}. There are strong
similarities between the vacuum evolution as a function of resolution 
scale and as a function of temperature. 
Both show the transition from partons to constituent quarks which has
also been phenomenologically checked in deep inelastic scattering at fixed
energy $\sqrt{s}$ as a function of photon virtuality $Q^2$ see Ref. 
\cite{Dosch:1997nw}
and Fig. (2). Numerically, the transition temperature $2\pi T_c$ 
is about equal to the chiral symmetry breaking scale $k_{\chi s b}$
where the quarks 
condense in the vacuum. Going from the ultraviolet to the 
infra-red the strong attractive
quark-antiquark interaction leads to a condensation of quark pairs in the
vacuum. At the resolution $k_{\chi s b}$ the effective mass parameter 
of the mesons equals zero. The critical exponents agree with the
$0(4)$ critical exponents, but differ from the mean field values \cite{SBW}.

In large $N_c$ approximation we have shown that the renormalization group 
flow equations are identical to the gap equations of the NJL model
\cite{Meyer:2001zp}. 
Further it is
possible to calculate the eigenmode spectrum of the Euclidean Dirac
operator \cite{Spitzenberg:2002tq}
which previously was only known for very small eigenvalues in random matrix
theories. 

The full equations 
possess extra complexity via the expansion of all 
running couplings and wavefunction renormalization parameters 
in $\Phi^2$. As mentioned above the global form of the meson potential
$U(\Phi^2)$ changes from the parabolic shape in the ultraviolet to a 
mexican hat shape in the
infra-red. This evolution leads to a change of the minimum of the effective
potential with resolution $k$. We track all couplings in Taylor
series
around this running minimum with terms up to 5th order in $\Phi^2$. This 
leads to 20 coupled non-linear differential equations. Including meson 
loop terms the infra-red properties of the system depend only on the
mass term in the ultraviolet and no longer on the starting ultraviolet
cut-off.
We have a real fixed point behaviour for the pion decay constant, 
meson quark coupling, constituent quark mass and quark condensate.
In comparison to NJL the flow equations
have predictive power since only one mass scale fixes all the other
dimensionful parameters.

For finite density \cite{Meyer:1999bz} 
 the model has similar deficiences as the conventional
NJL model. It overbinds and restores the chiral symmetry too early. 
We have improved the model by including $\omega$-repulsion and plan to do
this  calculation including loop corrections.

The very successful chiral perturbation 
theory 
 alone cannot connect to QCD in the
ultraviolet. Typically it starts to fail when the resolution scale is of
the 
order of half 
the $\rho$ mass. Hybrid models like the one presented here
containing   meson degrees of
freedom {\em and} quarks extend to larger momentum 
scales, which is a definite advantage. There are, however, two problems:
Firstly, the
number of mesons in the theory should include 
also the vector mesons.  Secondly, the size of the scalar 
Yukawa couplings increases dramatically in the ultraviolet, 
therefore it seems difficult  
to cross over to
the asymptotically free gauge theory of QCD.
For nuclei a theory with quarks and nucleons \cite{Meyer:1999bx} seems to be
the most efficient way to describe the transition from purely
nucleonic matter to quark matter at high baryon density
\cite {Schwenzer}.
Our model like most of the other hybrid models 
lacks a field theoretic mechanism to avoid the appearance of free quarks
as 
asymptotic states. A major progress
would be to capture  the smooth infra-red limit of lattice QCD
in a field theoretic continuum picture. 

\vspace{1.0cm}
\noindent
{\bf Acknowledgements:} 
We thank B. J. Schaefer for his critical reading of the manuscript.

\markright{References}

\end{document}